\documentclass[runningheads]{llncs}
\usepackage{graphicx}
\usepackage{algorithmic}
\usepackage[ruled,linesnumbered]{algorithm2e}
\usepackage{textcomp}
\usepackage{hyperref}       
\usepackage{url}            
\usepackage{booktabs}       
\usepackage{nicefrac}
\usepackage{microtype}      
\usepackage{changes}
\usepackage{lipsum}
\usepackage{enumitem}
\usepackage{listings}
\usepackage{array}
\usepackage{subcaption}
\usepackage{wrapfig}
\usepackage{mathtools,cuted}
\usepackage{color}
\usepackage{colortbl}
\usepackage{cases}
\usepackage{amsmath}
\usepackage{float}

\usepackage{stackengine,graphicx}
\stackMath

\usepackage{amssymb}
\usepackage[normalem]{ulem}
\usepackage{bm}
\usepackage{mathtools}
\usepackage{mathrsfs}
\usepackage{verbatim}
\usepackage{comment}

\newcommand{\xbf}{\mathbf{x}}

\newcommand{\bbf}{\mathbf{b}}
\newcommand{\nbf}{\mathbf{n}}

\newcommand{\fbf}{\mathbf{f}}

\newcommand{\Fbf}{\mathbf{F}}
\newcommand{\Pbf}{\mathbf{P}}

\DeclareMathOperator{\prox}{\mathrm{prox}}

\newcommand{\CC}{\mathbb{C}}

\DeclareMathOperator*{\argmin}{arg\,min}

\newcommand{\V}{\mathcal{V}}

\newcommand{\D}{\mathcal{D}}

\newcommand{\LL}{\mathcal{L}}

\usepackage{tikz}
\usetikzlibrary{shapes,arrows,chains}

\begin{document}

\title{Multi-task Magnetic Resonance Imaging 
Reconstruction using Meta-learning \thanks{This work is supported by the National Institute of Health under grants R21EB031185, R01AR081344, R01AR079442, and R56AR081017.}}
\titlerunning{}
\author{Wanyu Bian\inst{1} \and 
	Albert Jang\inst{1} \and
	Fang Liu\inst{1}}
%
\authorrunning{W. Bian et al.}

\institute{Harvard Medical School, Boston, MA 02115, USA\\
	\email{ \{bian4, awjang, fliu12\}@mgh.harvard.edu}}

\maketitle              

\begin{abstract}
Using single-task deep learning methods to reconstruct Magnetic Resonance Imaging (MRI) data acquired with different imaging sequences is inherently challenging. The trained deep learning model typically lacks generalizability, and the dissimilarity among image datasets with different types of contrast leads to suboptimal learning performance. This paper proposes a meta-learning approach to efficiently learn image features from multiple MR image datasets. Our algorithm can perform multi-task learning to simultaneously reconstruct MR images acquired using different imaging sequences with different image contrasts. The experiment results demonstrate the ability of our new meta-learning reconstruction method to successfully reconstruct highly-undersampled k-space data from multiple MRI datasets simultaneously, outperforming other compelling reconstruction methods previously developed for single-task learning.

\end{abstract}
\section{Introduction}
MR images acquired using different imaging sequences can provide complementary soft tissue contrasts, which are collectively used for clinical disease diagnosis. However, MRI is a slow imaging modality, resulting in a high sensitivity to subject motion. Long acquisition time is also undesirable, as it leads to lower patient throughput than other popular imaging modalities. Accelerated MRI acquisition and reconstruction are highly desirable and remain an active research topic \cite{sodickson1997simultaneous,griswold2002generalized,pruessmann1999sense,lustig2007sparse,lustig2010spirit,otazo2010combination,chen2021learnable}. In the past years, deep learning methods \cite{knoll2020deep} have shown great potential to enable rapid MRI. Deep learning approaches typically use task-specific deep networks to learn image features associated with incomplete k-space sampling and aim to remove image artifacts and noises caused by k-space undersampling during accelerated acquisition. While the conventional single-task models can work well when the training and testing data stem from the same data distribution (e.g., acquired using the same imaging sequence), they typically underperform when training and testing data differ substantially (e.g., acquired from different imaging sequences) \cite{antun2020instabilities,shimron2022implicit,liu2019santis}. Therefore, multi-task deep learning using methods to synergistically learn image features across multiple image datasets with different image contrasts is highly desirable for robust and high-efficient reconstruction.

MR images acquired using different imaging sequences can provide complementary soft tissue contrasts, which are collectively used for clinical disease diagnosis. However, MRI is a slow imaging modality due to its inherent sequential acquisition mode, resulting in a high sensitivity to subject motion. Long acquisition time is also undesirable, as it leads to lower patient throughput than other popular imaging modalities such as Computed Tomography (CT) and Ultrasound. Accelerated MRI acquisition and reconstruction are highly desirable and remain an active research topic in the MRI research community \cite{sodickson1997simultaneous,griswold2002generalized,pruessmann1999sense,lustig2007sparse,lustig2010spirit,otazo2010combination,bian2021optimization,bian2022learnable}. In the past years, deep learning methods \cite{knoll2020deep} have shown great potential to enable rapid MRI. Deep learning approaches typically use task-specific deep networks to learn image features associated with incomplete k-space sampling and aim to remove image artifacts and noises caused by k-space undersampling during accelerated acquisition. The trained models can then be applied to reconstruct newly undersampled data in the testing phase. While these single-task models can work well when the training and testing data stem from the same data distribution (e.g., acquired using the same imaging sequence), they typically underperform when training and testing data differ substantially (e.g., acquired from different imaging sequences) \cite{antun2020instabilities,shimron2022implicit,liu2019santis}. Therefore, multi-task deep learning using methods to synergistically learn image features across multiple image datasets with different image contrasts is highly desirable for robust and high-efficient reconstruction. 

This paper proposes a meta-learning framework to learn image features from multiple MR image datasets. Meta-learning is a stacking ensemble learning \cite{dvzeroski2004combining} scheme that is considered a process of ``learning-to-learn", which can learn to improve the learning algorithm and parameter generalizability over multiple training episodes, thus enabling each task to learn better \cite{hospedales2021meta}. In this study, we developed a bi-level meta-learning reconstruction framework (i.e., including base-level and meta-level) to handle the learning of multiple image datasets. At the base-level, we introduce new deep networks (e.g., base-learners) by unrolling proximal gradient descent in both image and k-space domains to cross-learn the image and frequency domain features for single image contrast. In the meta-level, we introduce an optimization algorithm that can alternatively optimize the base-learners and one additional meta-learner for efficiently characterizing mutual correlation among multiple image datasets. Our bi-level meta-learning can simultaneously reconstruct highly-undersampled k-space data acquired using different imaging sequences with an optimal reconstruction for all image contrasts. Meta 

Our contribution can be summarized as follows:\\
1.	A novel bi-level meta-learning framework is proposed to reconstruct highly-undersampled MRI datasets acquired using different imaging sequences. The trained meta-learning model can achieve improved reconstruction performance superior to standard deep learning methods, which can only perform well on single-task reconstruction.\\
2.	An unrolled network is proposed to extend proximal gradient descent  \cite{parikh2014proximal} on both image and k-space domains, enabling cross-domain learning of MRI data and leading to a superior reconstruction performance for every single contrast.\\
3.	The proposed algorithm is validated for reconstructing a set of knee MRIs acquired at different imaging contrasts and planes. With the testing at various acceleration rates, the proposed meta-learning demonstrates superb reconstruction performance through faithfully removing image artifacts and noises, preserving truthful image textures and details, and maintaining high image sharpness and conspicuity.

\section{Methodology and Algorithm}

Considering the multi-coil MR images at multi-task image datasets, we denote each image as  $\xbf_{i,j} \in \CC^n$ for $i=1, \cdots, m$ and $j = 1,\cdots, c$, where $m$ is the number of tasks (i.e., the acquired image sequences), $c$ is the number of receiving coils and $n$ is the number of pixels in the image. The corresponding undersampled k-space data can be formulated as  $ \fbf_{i,j} = \Pbf \Fbf \xbf_{i,j} + \nbf_{i,j}$, where $\Fbf$ denotes the MR encoding matrix, $\Pbf$ is a binary undersampling matrix, and  $\nbf_{i,j}$  is the noise. We aim to handle this reconstruction by solving the following optimization problem using meta-learning:
\begin{subequations}\label{model}
\begin{align}
\min_{\xbf_{i,j}} \,\, & \Psi_{\Theta,W}(\xbf_{i,j})  := 
\textstyle{\sum\limits_{i = 1}^{m} \sum\limits_{j = 1}^{c}  \frac{1}{2} \| \Pbf \Fbf \xbf_{i,j}  - \fbf_{i,j}\|_2^2   + G_{w_i} (\xbf_{i,j}) + K_{w_i} (\Fbf(\xbf_{i,j})) }, \label{model_up}\\
\mbox{s.t.}  \quad   & \xbf_{i,j} = \argmin_{\xbf_{i,j}} \textstyle{ \frac{1}{2} \| Z_{w_i} ( H_{\Theta}( [ J_{w_i}( \{\xbf_{j}\})_i ]_{i=1}^m) ) - RSS(\{\xbf_{j}\})_i \|^2_2 }. \label{model_lo}
\end{align}
\end{subequations}
In model \eqref{model}, $G, K, J, Z$ and $H$ all represent deep neural networks. We denote a collection of task-specific features as $W = \{w_i:i\in[m]\} $, which are generated by  \emph{base-learner}s $G, K, J, Z$ to learn from the individual task. The \emph{meta-learner} $H$, however, attempts to learn meta-knowledge  $\Theta$ that captures multi-task features from multiple datasets. Note that $W$ and $\Theta$ are network weights to be learned during training. 

More specifically, in the upper-level minimization problem \eqref{model_up}, two learnable regularization terms, namely $G_{w_i}$  and $K_{w_i}$, are applied to each task to learn image domain features and frequency domain (k-space) features, respectively, from the training data. The learner $G_{w_i}$ is designed to reduce image domain artifacts and noise while the learner $K_{w_i}$ intends to refine k-space to emphasize structural details, patterns, and textures. The lower-level optimization problem \eqref{model_lo} enforces data consistency to ensure that the output generated by the three learners $J, H, Z$ are close to the root sum-of-squares ($RSS$) of the multi-coil images. Each learner performed fundamentally different but essential roles in the learning process.

$J$ is the coil combination network that integrates the multi-coil images to produce a coil-combined image \cite{bian2022optimal}. Each task is associated with a task-specific $J$, and the resulting coil combined image $J_{w_i}(\{\xbf_{j}\})$ for $i$-th task is input into the high-dimensional meta-learner $H$ to extract meta-knowledge $\Theta$. The meta-knowledge is generated by learning the mapping $H_\Theta: \CC^{m n} \rightarrow \CC^{d n}$, where $d$ is the feature dimension (e.g., the number of kernels in the last convolutional layer of $H$). The meta-learner $H$ plays an essential role in generating, balancing, and compensating cross-correlations between the features of the different tasks. More importantly, the meta-knowledge $\Theta$ provides the parameter initialization of the entire model, which subsequently supports the individual base-learners to train their target tasks. The \emph{meta-distributor} $Z_{w_i}: \CC^{d n} \rightarrow \CC^{n}$ learns the mapping from the meta-knowledge $\Theta$ to the coil-combined image through $RSS$. More specifically, $Z_{w_i} ( H_\Theta ( [J_{w_1}(\{\xbf_{1,j}\}), \cdots, J_{w_m}(\{\xbf_{m,j}\})] ) ) $ attempts to output an image matching the coil-combined image $RSS(  \{ \xbf_{j} \}_{j=1}^c )_i$  for the $i$-th task. The meta-distributor maintains data consistency by learning the ability to distribute the meta-knowledge into each image. This enables knowledge sharing across multi-tasks, where the meta-knowledge can be efficiently distributed to each task according to their respective feature importance, thus improving the reconstruction quality of each task.

\begin{figure}
\includegraphics[width=1\linewidth]{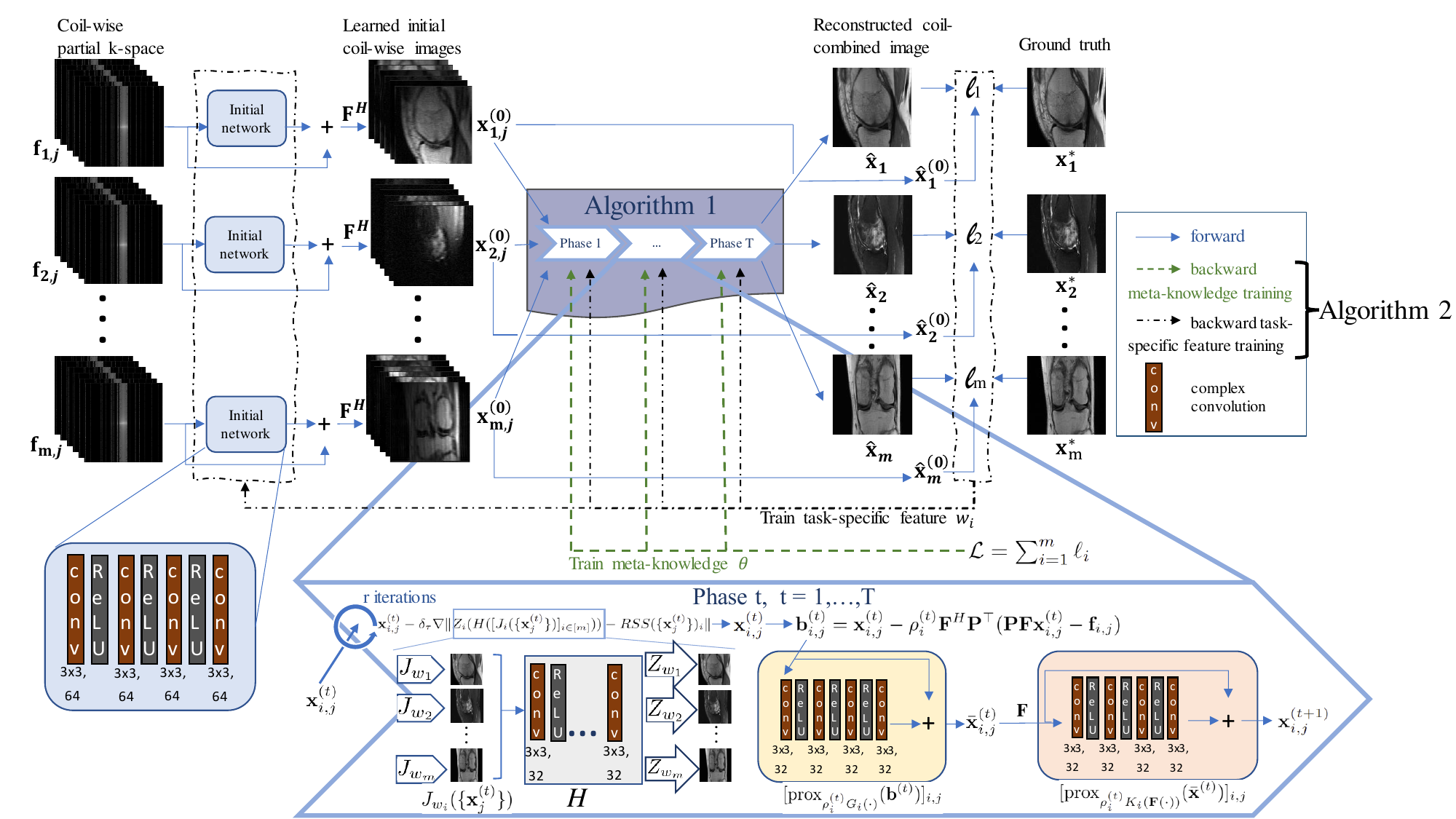}
\caption{The proposed multi-task meta-learning MRI reconstruction.}\label{fig:framework}
\end{figure}

In our algorithm, we developed a neural network architecture Fig. \ref{fig:framework} to implement training of the multi-task MRI reconstruction model. Our implementation consists of a forward step to unroll the optimization procedure through proximal gradient descent, and a backward meta-training to update weights and features in base-learners and meta-learner.

\subsection{Forward Learnable Optimization using Gradient Descent}
We propose to develop an unrolled network to solve both the upper-level variational minimization problem and the lower-level constrained optimization problem in model \eqref{model}. 
We introduce a proximal gradient descent \cite{parikh2014proximal} inspired algorithm \ref{alg:lda}, which can be parametrized as an iterative procedure with total $T$ iterations. Proximal gradient descent algorithm has been largely implemented in various applications for MRI reconstruction \cite{bian2020deep,bian2021optimization,bian2022optimization,bian2024improving,bian2023magnetic,chen2021variational,zhang2022extra}. The input $\xbf^{(0)}_{i,j}$ of the algorithm is obtained from an initial network that tries to pre-estimate the missing k-space data (similar to conventional parallel imaging methods such as GRAPPA \cite{griswold2002generalized}) therefore providing a good initial start for Algorithm \ref{alg:lda}. This helps reduce the number of iterations needed for convergence and decreases the overall computational complexity. The overall procedure consists of two for-loops: an outer and an inner loop, as follows:      
\begin{algorithm}[htbp]
\caption{Forward Learnable Descent Algorithm for solving \eqref{model}}
\label{alg:lda}
\begin{algorithmic}[1]
\STATE \textbf{Input:} $\xbf^{(0)}_{i,j},\delta_0, \rho_i^{(0)} $.
\FOR{$t=0,1,\dots,T-1$}
\FOR{$\tau =0,1,\dots,r$} \label{inner_s}
\STATE $ \xbf^{(t)}_{i,j}  \leftarrow {\xbf}^{(t)}_{i,j} - \delta_{\tau} \nabla \| Z_i ( H( [ J_i ( \{ {\xbf}_{j}^{(t)}\} )  ]_{i\in [m]})) - RSS( \{ \xbf^{(t)}_j \})_i \|$ \label{eq:initial}
\ENDFOR \label{inner_e}
\STATE $\bbf_{i,j}^{(t)} = \xbf_{i,j}^{(t)} - \rho^{(t)}_{i}  \Fbf^{H}  \Pbf^{\top}  (\Pbf \Fbf \xbf_{i,j}^{(t)} - \fbf_{i,j}),   \quad i = 1,\cdots, m, \, j =  1,\cdots, c,$ \label{eq:bi}
\STATE $\bar{\xbf}_{i,j}^{(t)}  = [\prox_{\rho^{(t)}_{i} G_i(\cdot) } ({\bbf}^{(t)})]_{i,j},  \quad i = 1,\cdots, m, \, j =  1,\cdots, c,$\label{eq:xbar}
\STATE ${\xbf}_{i,j}^{(t+1)} = [\prox_{ \rho^{(t)}_{i} K_i(\Fbf(\cdot))}  (\bar{\xbf}^{(t)})]_{i,j}, \quad i = 1,\cdots, m, \, j =  1,\cdots, c, $\label{eq:xij}
\ENDFOR{ and \textbf{output} $ \xbf^{(T)}_{i,j}$, $ Z_i ( H( [ J_i ( \{ {\xbf}_{j}^{(T)}\} )  ]_{i\in [m]}))$,  $ Z_i ( H( [ J_i ( \{ {\xbf}_{j}^{(0)} \} ) ]_{i\in [m]}))$ }\label{lda_end}
\end{algorithmic}
\end{algorithm}

The inner loop aims to solve the lower-level optimization in \eqref{model_lo}. We first calculate the gradient of the constrained least squares in \eqref{model_lo} and then iterate $r$ steps using gradient descent with a learnable step size $\delta_{\tau}$ shown in steps \ref{inner_s}-\ref{inner_e} in Algorithm \ref{alg:lda}.
In the outer loop, the updated ${\xbf}^{(t)}_{i,j}$ serves as the initial input for steps \ref{eq:bi}-\ref{eq:xij} of proximal gradient descent. The proximal operator is defined as $\prox_{\alpha R}(\bbf) := \argmin_{\xbf} \left\{\frac{1}{2 \alpha} \| \xbf - \bbf \|^2 + R(\xbf) \right\}$. In step \ref{eq:xbar} and \ref{eq:xij}, two proximal operators are conducted by alternating between image and k-space domains to ensure learning of both image and frequency features. Because the proximal operators do not have a closed-form solution for explicit implementation, we parametrize $\prox_{\rho^{(t)}_{i} G_i(\cdot) }$ and $\prox_{ \rho^{(t)}_{i} K_i(\Fbf(\cdot))} $ as two learnable deep residual networks in image and k-space domains, respectively, to emphasize artifacts and noise removal (in image domain) and structure preservation (in frequency domain).  Our final outputs in step \ref{lda_end} from Algorithm \ref{alg:lda} are reconstructed multi-coil images of all tasks $ \xbf^{(T)}_{i,j}$, we then apply $ Z_i, H, J_i$ to obtain their corresponding coil-combined, meta-knowledge distributed image $\hat{\xbf}_i = Z_i ( H( [ J_i ( \{ {\xbf}_{j}^{(T)}\} )  ]_{i\in [m]}))$. In addition, we also apply  $ Z_i, H, J_i$ to the initial reconstruction  $\xbf^{(0)}_{i,j}$ and get $ \hat{\xbf}_i^{(0)} = Z_i ( H( [J_i ( \{ {\xbf}_{j}^{(0)}) \} ]_{i\in [m]} ))$. Both $\hat{\xbf}_i$ and $\hat{\xbf}_i^{(0)}$ are then input to the loss function. Our overall loss function is designed as follows for the forward optimization:
\begin{subequations}
\label{loss_function}
\begin{align}
 & \textstyle{\LL(\Theta, W;  \D)  := \sum^{m}_{i=1} \ell_{i}( \Theta , w_i; \D_{i}) },  \text{ where each $\ell_{i}$ is called \emph{base-loss}:}\\
  &  \textstyle{\ell_{i}( \Theta , w_i; \D_{i}) = \| \hat{\xbf}_i(\Theta , w_i) - \xbf^*_i \|  + \lambda \|  \hat{\xbf}_i^{(0)}(\Theta , w_i)  - \xbf^*_i \| -\mu \text{SSIM}(\hat{\xbf}_i(\Theta , w_i), \xbf^*_i)  },
\end{align}
\end{subequations}
where we use $* $ to reflect the fully-sampled k-space and the ground truth for each task is $\xbf^*_i = RSS(\{ \xbf^*_j \}_{j=1}^c)_i $  and  $\D_{i} = {(\fbf_i,\xbf^*_i )}$  is the training data pair for the $i$-th task. The objective is to learn network weights $W$ and $\Theta$ during training. The loss function consists of three terms in \eqref{loss_function}. The first term is a standard supervised learning term forcing the final reconstructed image to approximate the target image from a fully-sampled k-space. The second term aims to learn a favorable initial input $\xbf^{(0)}_{i,j}$ for Algorithm \ref{alg:lda}. The third term uses the structural similarity index (SSIM) \cite{wang2004image} to enforce the similarity between the reconstructed and target images. Parameters $\lambda$ and $\mu$ are prescribed to balance the three terms.
\subsection{Backward Network Update through Meta-training}
The backward operation \eqref{eq:bi-level} is conducted during training to minimize the overall loss function in \eqref{loss_function} with respect to the network weights $\Theta$ and $W$. 
\begin{subequations}
\label{eq:bi-level}
\begin{align}
& \textstyle{\min_{ \Theta(W) } \LL(\Theta , W; \D)} \label{meta_up}\\
 \mbox{s.t.} \quad  &   w_i  = \textstyle{ \argmin_{w_i} \ell_i (\Theta (W), w_i; \D_{i}), \quad  i= 1, \cdots,m } \label{meta_lo}
\end{align}
\end{subequations}
The overall procedure to solve \eqref{eq:bi-level} is inspired by Model-Agnostic Meta-Learning (MAML) \cite{finn2017model} that consists of two for-loops: an outer and an inner loop, as follows:
\begin{algorithm}
\caption{Backward Meta-Training algorithm for solving \eqref{eq:bi-level}.}
\label{alg:alternating}
\begin{algorithmic}[1]
\STATE \textbf{Input:  $\D_i$ } 
\textbf{and initialize}  $ \Theta$, $ {w_i}$, $i=1,\cdots, m $.\\
\FOR{$l = 0,1,\dots, L$}
\STATE Sample training volume $ \V_i \subset \D_i$.
\FOR{$k=0,1,\dots,K$ (inner loop)} \label{meta_inner_s}
\STATE $ \Theta \leftarrow \Theta - \beta \nabla_{\Theta}\LL( \Theta, W ; \{ \V_i:i\in[m] \})$
\ENDFOR  \label{meta_inner_e}
\STATE $ w_i \leftarrow w_i - \alpha_{i} \nabla_{w_i}\ell_i ( \Theta, w_i ; \V_i), \quad  i=1,\cdots, m$
\ENDFOR{ and  \textbf{output} $\Theta$, $W = \{ w_1, \cdots, w_m \} $}
\end{algorithmic}
\end{algorithm}
We train the outer loop with a total of $L$ epochs. A mini-batch training with a randomly selected 3D image volume is used in each epoch $l \in \{0, \cdots, L\}$. The inner loop aims to learn meta-knowledge $\Theta(W)$ through the upper-level (meta-level) minimization in \eqref{meta_up}, where the meta-knowledge $\Theta$ is a function dependent on network weights $W$ (a.k.a., learning-to-learn). The upper-level minimization fixes $W$ and only updates $\Theta$ in the inner loop steps \ref{meta_inner_s}-\ref{meta_inner_e}, and then the updated $\Theta(W)$ is used as input into lower-level (base-level) minimization to optimize each $w_i$ in \eqref{meta_lo} in the outer loop. $\beta$ represents the meta-learning rate in the inner loop and $\alpha_i$ represents each base learning rate in the outer loop.
\section{Experiments}
\subsection{Experiment Setup}

The study was approved by the local IRB. Knee datasets with fully-sampled k-space were acquired on $25$ subjects (20/5 for training/testing). The datasets were acquired on a 3.0T scanner  with an 18-element knee coil array using four two-dimensional fast spin-echo (FSE) sequences, including coronal proton density-weighted FSE (Cor-PD) and T2-weighted FSE (Cor-T2), and sagittal proton density-weighted FSE (Sag-PD) and T2-weighted FSE (Sag-T2).


Experiments were performed to evaluate the efficacy and efficiency of our proposed method to reconstruct highly-undersampled multi-task knee MRI. K-space data were retrospectively undersampled \cite{lustig2007sparse} to simulate acceleration rates (AR) of $\{4, 5, 6\}$. To illustrate the effect of meta-learning, we denote our proposed method without the meta-learner $H_\Theta$ (i.e., only iterate steps \ref{eq:bi} to \ref{eq:xij} in Algorithm \ref{alg:lda} for m tasks separately) as single-task learning (STL) and the full version as multi-task meta-learning (MTML). All hyper-parameter selection is shown in the Appendix. We also compared our proposed STL and MTML against two recently developed reconstruction methods ISTA-Net \cite{zhang2018ista} and pMRI-Net \cite{bian2020deep}, both of which were previously shown to provide great performance in reconstructing undersampled single-task MRI data.
%
%
\subsection{Experimental Results}
\begin{table}[t]
\centering
\caption{Quantitative comparison of different methods. (ITL: proposed single-task learning; MTML: proposed multi-task meta-learning; PSNR: peak signal-to-noise ratio; NMSE: normalized mean squared error)} \label{Quant_table}
\begin{tabular}{c|ccc}
\hline
 &   & \textbf{Sag-T2} & \\
 AR  & 4x & 5x & 6x  \\
\hline
  &  PSNR/SSIM/NMSE  & PSNR/SSIM/NMSE  &  PSNR/SSIM/NMSE     \\
ISTA-Net\cite{zhang2018ista} & 34.6138/0.9687/0.1444 &  33.4308/0.9600/0.1659  &  31.5700/0.9458/0.2061  \\
pMRI-Net\cite{bian2020deep} & 36.1084/0.9769/0.1214 &  35.0315/0.9712/0.1375  &  34.3608/0.9672/0.1486  \\
ITL    & 37.4666/0.9825/0.1039 &  36.6242/0.9793/0.1144  &  35.9206/0.9763/0.1239  \\
MTML   & 38.3446/0.9854/0.0941 &  37.5966/0.9832/0.1023  &  36.8210/0.9799/0.1118  \\ 
\end{tabular}
\begin{tabular}{c|ccc}
\hline
 &   & \textbf{Cor-T2}  & \\
\hline
  &  PSNR/SSIM/NMSE  & PSNR/SSIM/NMSE & PSNR/SSIM/NMSE   \\
ISTA-Net\cite{zhang2018ista} & 34.1552/0.9628/0.1142 &  32.7243/0.9472/0.1382  &  30.0678/0.9155/0.1864  \\
pMRI-Net\cite{bian2020deep}  & 36.3567/0.9767/0.0867 &  35.5938/0.9728/0.0942  &  34.1399/0.9646/0.1108  \\
ITL & 37.0114/0.9796/0.0808 &  36.1727/0.9758/0.0883  &  35.7136/0.9739/0.0930  \\
MTML   & 37.8478/0.9828/0.0740 &  37.1512/0.9812/0.0793  &  36.6107/0.9781/0.0843  \\
\end{tabular}
\begin{tabular}{c|ccc}
\hline
 &  & \textbf{Sag-PD} &  \\
\hline
  &  PSNR/SSIM/NMSE  & PSNR/SSIM/NMSE & PSNR/SSIM/NMSE   \\
ISTA-Net\cite{zhang2018ista} & 33.1395/0.9541/0.0774 & 31.9699/0.9495/0.0877  & 29.4236/0.9280/0.1176 \\
pMRI-Net\cite{bian2020deep} & 35.0189/0.9685/0.0623 & 34.1714/0.9588/0.0690  & 31.8230/0.9365/0.0906 \\
ITL & 35.9640/0.9736/0.0558 & 35.0720/0.9680/0.0617  & 34.0919/0.9619/0.0691 \\
MTML   & 37.0246/0.9784/0.0493 & 36.2652/0.9754/0.0537  & 35.1452/0.9688/0.0611 \\ 
\end{tabular}
\begin{tabular}{c|ccc}
\hline
 &  & \textbf{Cor-PD } \\
\hline
  &  PSNR/SSIM/NMSE  & PSNR/SSIM/NMSE & PSNR/SSIM/NMSE   \\
ISTA-Net\cite{zhang2018ista} & 32.3933/0.9640/0.1030 & 31.1878/0.9555/0.1150  &  29.1795/0.9603/0.1281 \\
pMRI-Net\cite{bian2020deep} & 33.5534/0.9695/0.0937 & 32.7408/0.9667/0.1003  &  30.8576/0.9560/0.1178  \\
ITL & 35.5481/0.9760/0.0809 & 33.5579/0.9698/0.0939  &  31.9874/0.9625/0.1045 \\
MTML   & 36.6621/0.9797/0.0748 & 35.4338/0.9765/0.0817  &  32.8284/0.9663/0.0999 \\ 
\hline
\end{tabular}
\end{table}
\begin{table}
\centering
\caption{The hyper-parameter selection. (All the methods were implemented using PyTorch, and training and testing were conducted using two NVIDIA A100 GPUs.) }\label{parameter}
\label{hyperparameters}
\begin{tabular}{|c|c|c|c|c|c|c|c|}
\hline
Initializer & Xavier \cite{glorot2010understanding} & 
Learning rate for $\Theta$ & $\beta=1e-4$ & $L$ & 100 & $K$ & $10$\\ 
\hline
Optimizer & ADAM \cite{kingma2014adam} & Learning rate for PD & $\alpha_i = 5e-4$  & $T$ & $5$ & $r$ & $5$\\
\hline
Training data & 20 subjects & Learning rate for T2  & $\alpha_i = 2e-4$  & $\rho_i^{(0)}$ & $0.5$ & $\delta_0$ & $0.5$ \\
\hline
Testing data & 5 subjects & Kernel size  & $3\times 3 \times 32$ & $\lambda$ & $1e-4$ & $\mu$ & $1$ \\
\hline
 Image size & $300 \times 300$ & Slices & 30 to 38 & $m$ & $4$ &  $c$ & $18$\\
\hline
\end{tabular}
\end{table}
\begin{figure}
\includegraphics[width=0.03\linewidth]{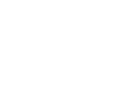}
\includegraphics[width=1\linewidth]{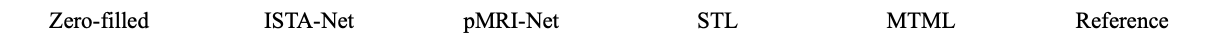}\\
\includegraphics[width=0.03\linewidth]{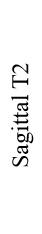}
\includegraphics[width=1\linewidth]{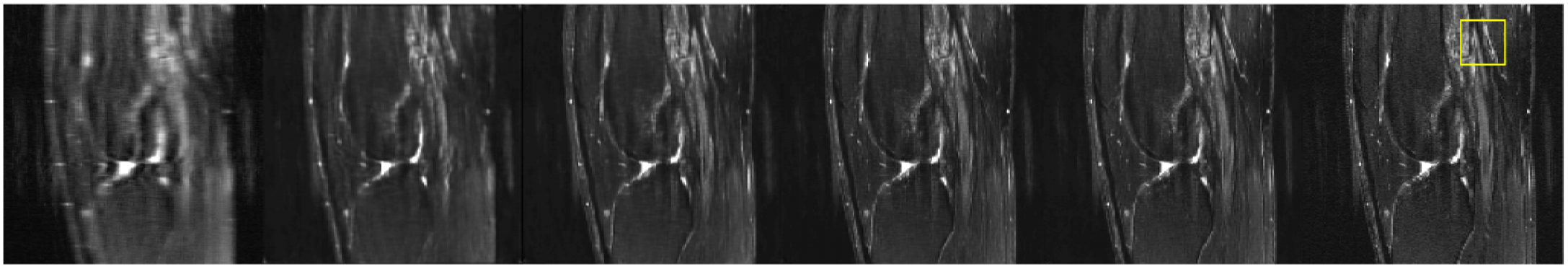}\\
\includegraphics[width=0.03\linewidth]{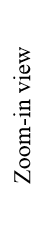}
\includegraphics[width=1\linewidth]{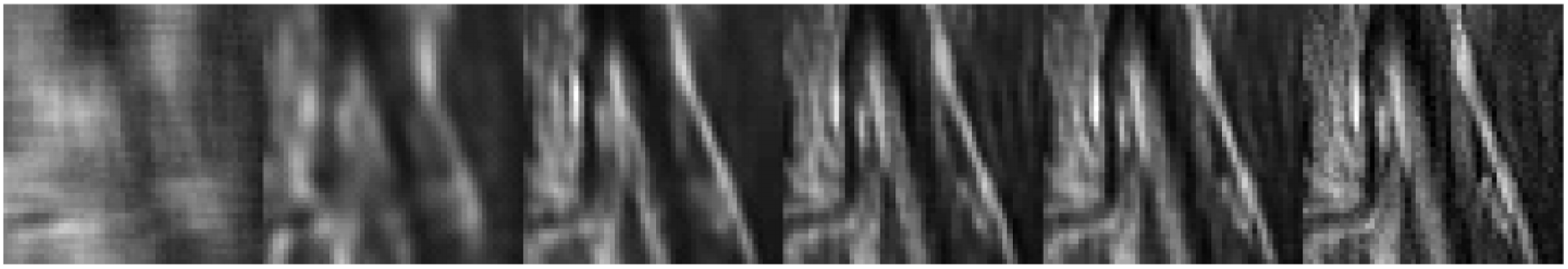}\\
\includegraphics[width=0.03\linewidth]{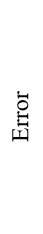}
\includegraphics[width=1.057\linewidth]{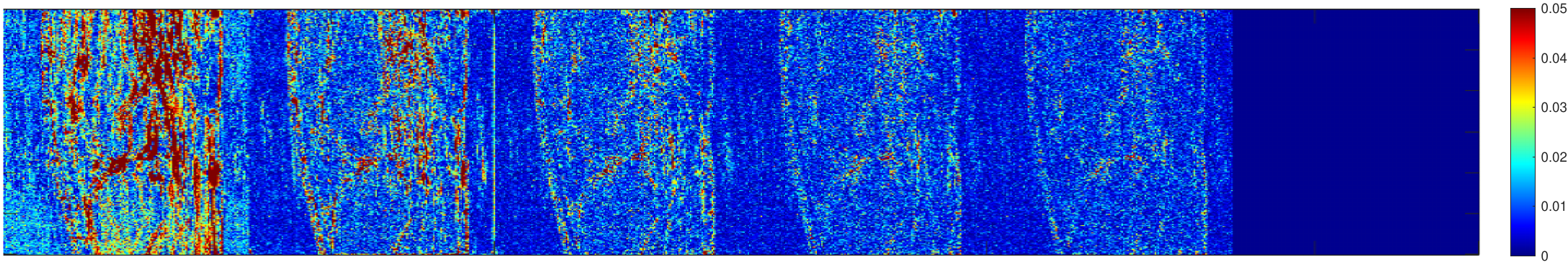}\\
\includegraphics[width=0.03\linewidth]{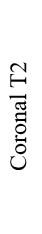}
\includegraphics[width=1\linewidth]{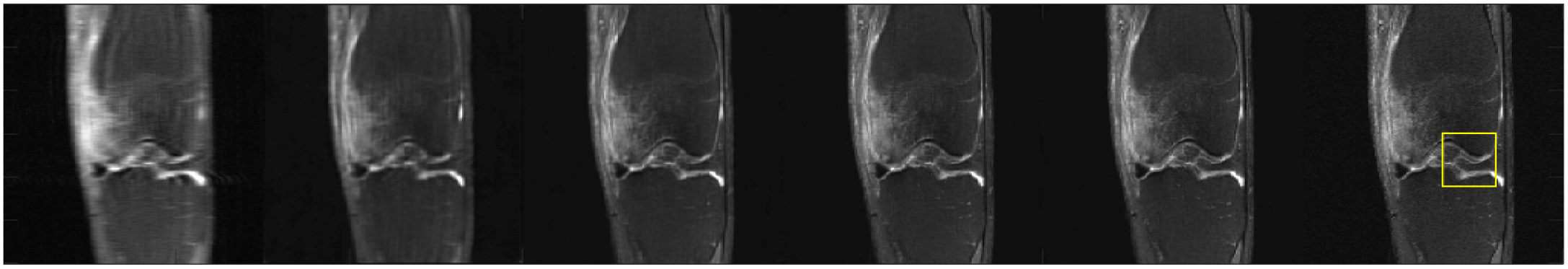}\\
\includegraphics[width=0.03\linewidth]{figure/zoom.png}
\includegraphics[width=1\linewidth]{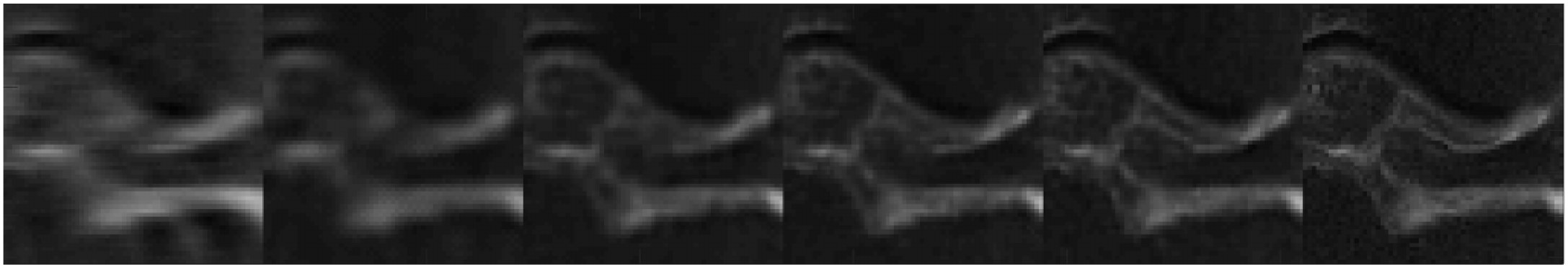}\\
\includegraphics[width=0.03\linewidth]{figure/error.png}
\includegraphics[width=1.057\linewidth]{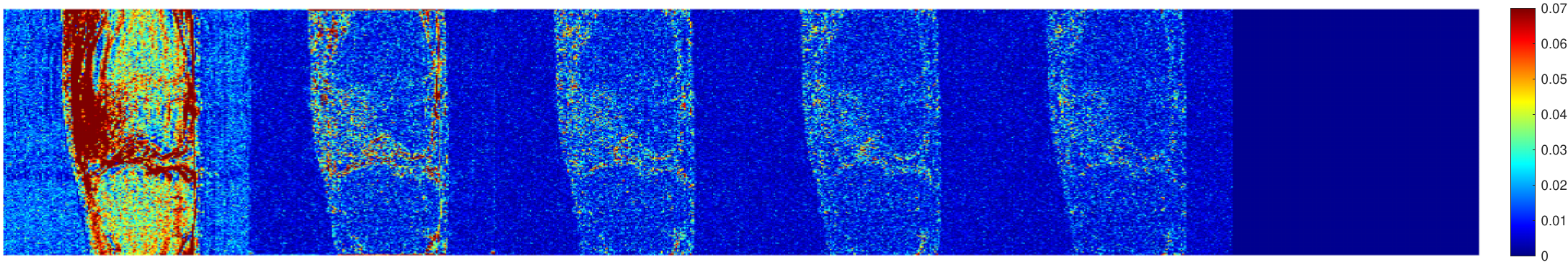}
\caption{Comparison of  Sag-T2 and Cor-T2 contrasts at AR = 6.}\label{fig:t2}
\end{figure}

Fig. \ref{fig:t2} compares the reconstructed images for Sag-T2 and Cor-T2 contrasts from different methods at AR=6. While ISTA-Net  and pMRI-Net can largely remove the aliasing artifacts at this undersampling level, both methods yielded suboptimal reconstruction performance with noticeable image blurring in the reconstructed images. Our proposed STL can enable better reconstruction than ISTA-Net and pMRI-Net with improved image sharpness, due to its ability to cross-learn both image and k-space features, therefore better removing noises and artifacts and meanwhile preserving high-frequency image features. Our proposed MTML using meta-learning performs the best, outperforming all single-task learning methods. The pixel-wise error maps show that MTML has the smallest reconstruction errors compared to the fully-sampled reference, and the error is more homogeneous across the entire image domain for MTML. The zoom-in images further highlight the superb reconstruction performance of MTML where the tissue boundaries are better characterized at all tissue types (cartilage, meniscus, and muscle), and tissue texture, sharpness, and conspicuity are well-preserved. Similar comparisons of reconstructed images for the Sag-PD and Cor-PD sequences are shown in the Table \ref{hyperparameters}, where MTML also consistently enables the best and most balanced performance on different types of contrast. 

The qualitative comparison in Fig. \ref{fig:t2} is further verified by different quantitative metrics shown in Table 1.  The table shows the mean PSNR, SSIM, and NMSE for all tested subjects for all methods at three ARs=$\{4, 5, 6\}$. MTML achieves the best quantitative performance in all metrics at all ARs, consistent with the qualitative assessment in exemplified figures.

Our proof-of-concept study using multi-task meta-learning opens a new window for further investigating robust and efficient multi-contrast large-scale MRI reconstruction algorithms.

%

\bibliographystyle{splncs04}
\bibliography{reference}

\begin{thebibliography}{10}
\providecommand{\url}[1]{\texttt{#1}}
\providecommand{\urlprefix}{URL }
\providecommand{\doi}[1]{https://doi.org/#1}

\bibitem{antun2020instabilities}
Antun, V., Renna, F., Poon, C., Adcock, B., Hansen, A.C.: On instabilities of
  deep learning in image reconstruction and the potential costs of ai.
  Proceedings of the National Academy of Sciences  \textbf{117}(48),
  30088--30095 (2020)

\bibitem{bian2022optimization}
Bian, W.: Optimization-Based Deep learning methods for Magnetic Resonance
  Imaging Reconstruction and Synthesis. Ph.D. thesis, University of Florida
  (2022)

\bibitem{bian2020deep}
Bian, W., Chen, Y., Ye, X.: Deep parallel mri reconstruction network without
  coil sensitivities. In: Machine Learning for Medical Image Reconstruction:
  Third International Workshop, MLMIR 2020, Held in Conjunction with MICCAI
  2020, Lima, Peru, October 8, 2020, Proceedings 3. pp. 17--26. Springer (2020)

\bibitem{bian2022optimal}
Bian, W., Chen, Y., Ye, X.: An optimal control framework for joint-channel
  parallel mri reconstruction without coil sensitivities. Magnetic Resonance
  Imaging  (2022)

\bibitem{bian2021optimization}
Bian, W., Chen, Y., Ye, X., Zhang, Q.: An optimization-based meta-learning
  model for mri reconstruction with diverse dataset. Journal of Imaging
  \textbf{7}(11), ~231 (2021)

\bibitem{bian2023magnetic}
Bian, W., Jang, A., Liu, F.: Magnetic resonance parameter mapping using
  self-supervised deep learning with model reinforcement. ArXiv  (2023)

\bibitem{bian2024improving}
Bian, W., Jang, A., Liu, F.: Improving quantitative mri using self-supervised
  deep learning with model reinforcement: Demonstration for rapid t1 mapping.
  Magnetic Resonance in Medicine  (2024)

\bibitem{bian2022learnable}
Bian, W., Zhang, Q., Ye, X., Chen, Y.: A learnable variational model for joint
  multimodal mri reconstruction and synthesis. In: International Conference on
  Medical Image Computing and Computer-Assisted Intervention. pp. 354--364.
  Springer (2022)

\bibitem{chen2021learnable}
Chen, Y., Liu, H., Ye, X., Zhang, Q.: Learnable descent algorithm for nonsmooth
  nonconvex image reconstruction. SIAM Journal on Imaging Sciences
  \textbf{14}(4),  1532--1564 (2021)

\bibitem{chen2021variational}
Chen, Y., Ye, X., Zhang, Q.: Variational model-based deep neural networks for
  image reconstruction. Handbook of Mathematical Models and Algorithms in
  Computer Vision and Imaging: Mathematical Imaging and Vision pp. 1--29 (2021)

\bibitem{dvzeroski2004combining}
D{\v{z}}eroski, S., {\v{Z}}enko, B.: Is combining classifiers with stacking
  better than selecting the best one? Machine learning  \textbf{54},  255--273
  (2004)

\bibitem{finn2017model}
Finn, C., Abbeel, P., Levine, S.: Model-agnostic meta-learning for fast
  adaptation of deep networks. In: International conference on machine
  learning. pp. 1126--1135. PMLR (2017)

\bibitem{glorot2010understanding}
Glorot, X., Bengio, Y.: Understanding the difficulty of training deep
  feedforward neural networks. In: Proceedings of the thirteenth international
  conference on artificial intelligence and statistics. pp. 249--256. JMLR
  Workshop and Conference Proceedings (2010)

\bibitem{griswold2002generalized}
Griswold, M.A., et~al.: Generalized autocalibrating partially parallel
  acquisitions (grappa). Magnetic Resonance in Medicine: An Official Journal of
  the International Society for Magnetic Resonance in Medicine  \textbf{47}(6),
   1202--1210 (2002)

\bibitem{hospedales2021meta}
Hospedales, T., Antoniou, A., Micaelli, P., Storkey, A.: Meta-learning in
  neural networks: A survey. IEEE transactions on pattern analysis and machine
  intelligence  \textbf{44}(9),  5149--5169 (2021)

\bibitem{kingma2014adam}
Kingma, D.P., Ba, J.: Adam: {A} method for stochastic optimization. In: Bengio,
  Y., LeCun, Y. (eds.) 3rd International Conference on Learning
  Representations, {ICLR} 2015, San Diego, CA, USA, May 7-9, 2015, Conference
  Track Proceedings (2015)

\bibitem{knoll2020deep}
Knoll, F., Hammernik, K., Zhang, C., Moeller, S., Pock, T., Sodickson, D.K.,
  Akcakaya, M.: Deep-learning methods for parallel magnetic resonance imaging
  reconstruction: A survey of the current approaches, trends, and issues. IEEE
  signal processing magazine  \textbf{37}(1),  128--140 (2020)

\bibitem{liu2019santis}
Liu, F., Samsonov, A., Chen, L., Kijowski, R., Feng, L.: Santis:
  sampling-augmented neural network with incoherent structure for mr image
  reconstruction. Magnetic resonance in medicine  \textbf{82}(5),  1890--1904
  (2019)

\bibitem{lustig2007sparse}
Lustig, M., Donoho, D., Pauly, J.M.: Sparse mri: The application of compressed
  sensing for rapid mr imaging. Magnetic Resonance in Medicine: An Official
  Journal of the International Society for Magnetic Resonance in Medicine
  \textbf{58}(6),  1182--1195 (2007)

\bibitem{lustig2010spirit}
Lustig, M., Pauly, J.M.: Spirit: iterative self-consistent parallel imaging
  reconstruction from arbitrary k-space. Magnetic resonance in medicine
  \textbf{64}(2),  457--471 (2010)

\bibitem{otazo2010combination}
Otazo, R., Kim, D., Axel, L., Sodickson, D.K.: Combination of compressed
  sensing and parallel imaging for highly accelerated first-pass cardiac
  perfusion mri. Magnetic resonance in medicine  \textbf{64}(3),  767--776
  (2010)

\bibitem{parikh2014proximal}
Parikh, N., Boyd, S.: Proximal algorithms. Foundations and Trends in
  optimization  \textbf{1}(3),  127--239 (2014)

\bibitem{pruessmann1999sense}
Pruessmann, K.P., Weiger, M., Scheidegger, M.B., Boesiger, P.: Sense:
  sensitivity encoding for fast mri. Magnetic Resonance in Medicine: An
  Official Journal of the International Society for Magnetic Resonance in
  Medicine  \textbf{42}(5),  952--962 (1999)

\bibitem{shimron2022implicit}
Shimron, E., Tamir, J.I., Wang, K., Lustig, M.: Implicit data crimes: Machine
  learning bias arising from misuse of public data. Proceedings of the National
  Academy of Sciences  \textbf{119}(13),  e2117203119 (2022)

\bibitem{sodickson1997simultaneous}
Sodickson, D.K., Manning, W.J.: Simultaneous acquisition of spatial harmonics
  (smash): fast imaging with radiofrequency coil arrays. Magnetic resonance in
  medicine  \textbf{38}(4),  591--603 (1997)

\bibitem{wang2004image}
Wang, Z., et~al.: Image quality assessment: from error visibility to structural
  similarity. IEEE transactions on image processing  \textbf{13}(4),  600--612
  (2004)

\bibitem{zhang2018ista}
Zhang, J., Ghanem, B.: Ista-net: Interpretable optimization-inspired deep
  network for image compressive sensing. In: Proceedings of the IEEE conference
  on computer vision and pattern recognition. pp. 1828--1837 (2018)

\bibitem{zhang2022extra}
Zhang, Q., Ye, X., Chen, Y.: Extra proximal-gradient network with learned
  regularization for image compressive sensing reconstruction. Journal of
  Imaging  \textbf{8}(7), ~178 (2022)

\end{thebibliography}

\end{document}


\section{Appendix}
\begin{figure}
\includegraphics[width=0.03\linewidth]{figure/white.png}
\includegraphics[width=1\linewidth]{figure/methods.png}\\
\includegraphics[width=0.03\linewidth]{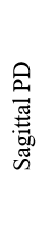}
\includegraphics[width=1\linewidth]{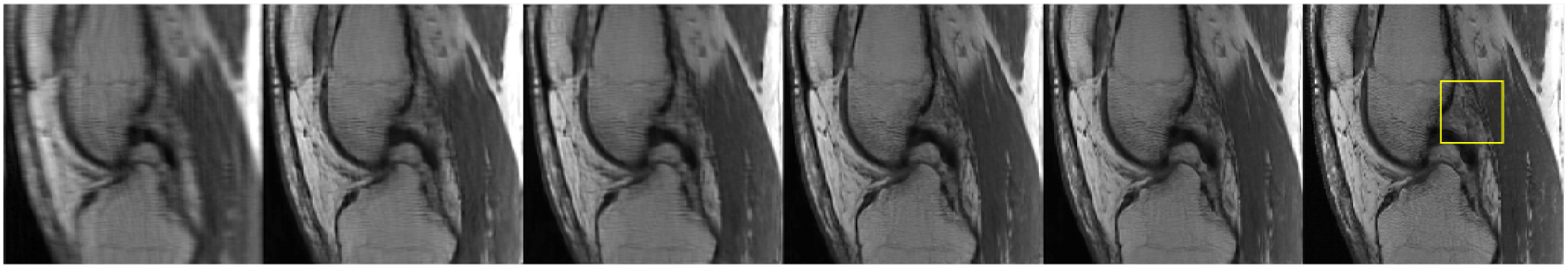}\\
\includegraphics[width=0.03\linewidth]{figure/zoom.png}
\includegraphics[width=1\linewidth]{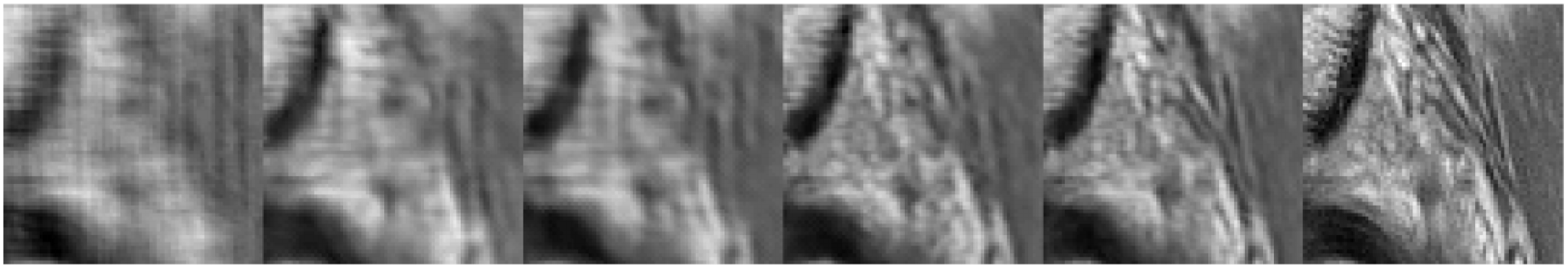}\\
\includegraphics[width=0.03\linewidth]{figure/error.png}
\includegraphics[width=1.057\linewidth]{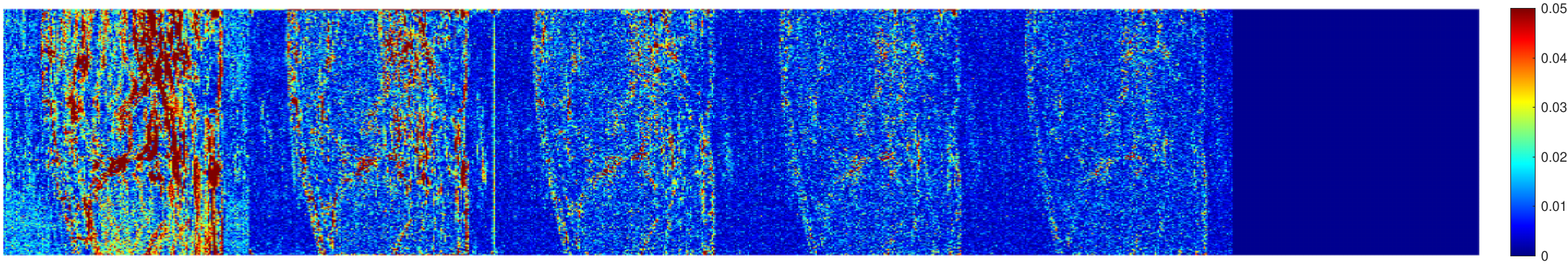}\\
\includegraphics[width=0.03\linewidth]{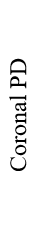}
\includegraphics[width=1\linewidth]{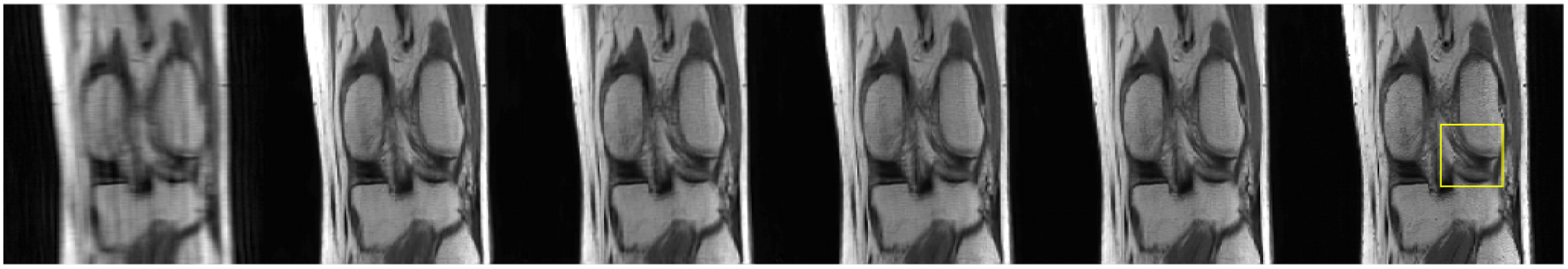}\\
\includegraphics[width=0.03\linewidth]{figure/zoom.png}
\includegraphics[width=1\linewidth]{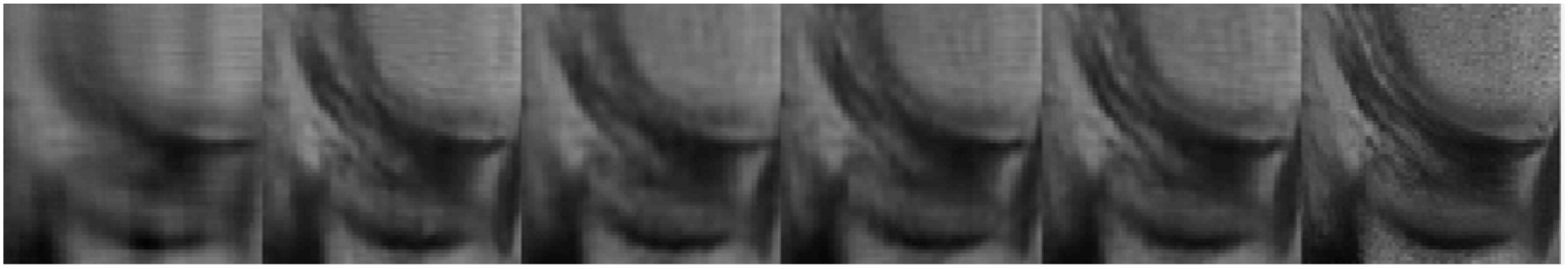}\\
\includegraphics[width=0.03\linewidth]{figure/error.png}
\includegraphics[width=1.057\linewidth]{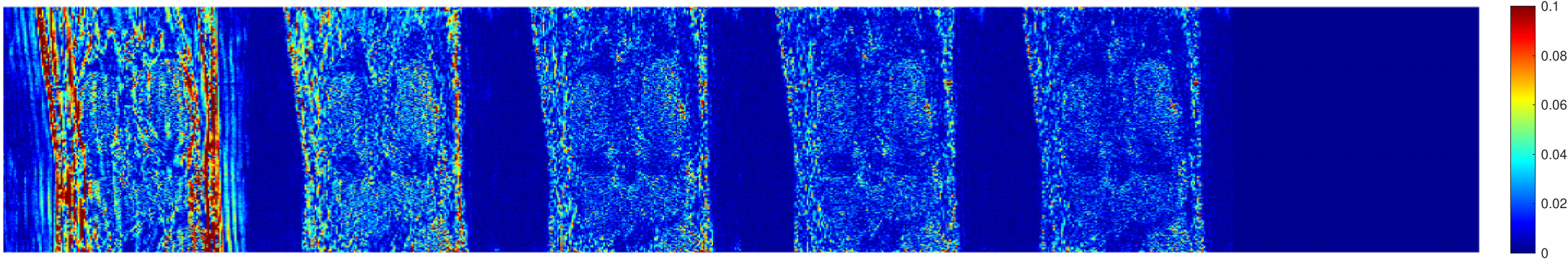}
\caption{Comparison of Sag-PD and Cor-PD contrasts at AR = 6. }
\label{fig:pd}
\end{figure}
\newpage

\bibliographystyle{splncs04}
\bibliography{reference}